\documentclass[aps,onecolumn,a4paper,showkeys,showpacs,prl,reprint,superscriptaddress]{revtex4-1}
\usepackage{amsmath}
\usepackage{amsfonts}
\usepackage{amssymb}
\usepackage{graphicx}
\usepackage{hyperref}
\hypersetup{citecolor=blue,filecolor=blue,linkcolor=blue,menucolor=blue,urlcolor=black,colorlinks=true}

\newcommand{\NTT}{\affiliation{NTT Basic Research Laboratories, NTT Corporation, 3-1 Morinosato-Wakamiya, Atsugi, Kanagawa, 243-0198, Japan}}

\newcommand{\Bin}{\mathbf{B}_{\parallel}}
\newcommand{\Bpp}{\mathbf{B}_{\perp}}

\newcommand{\red}[1]{\textcolor{black}{#1}}
\newcommand{\green}[1]{\textcolor{black}{#1}}
\newcommand{\blue}[1]{\textcolor{black}{#1}}

\begin{document}
\author{Hiraku~Toida}\email{toida.hiraku@lab.ntt.co.jp}\NTT
\author{Yuichiro~Matsuzaki}\NTT
\author{Kosuke~Kakuyanagi}\NTT
\author{Xiaobo~Zhu}\email{Current address: Synergetic Innovation Center of Quantum Information and Quantum Physics, University of Science and Technology of China, Hefei, Anhui 230026, China}\NTT
\author{William~J.~Munro}\NTT
\author{Hiroshi~Yamaguchi}\NTT
\author{Shiro~Saito}\NTT

\title{Electron paramagnetic resonance spectroscopy using a single artificial atom}

\maketitle

\textbf{
Electron paramagnetic resonance (EPR) spectroscopy is an important technology in physics, chemistry, materials science, and biology \cite{Schweiger2001}.
Sensitive detection with a small sample volume is a key objective in these areas, because it is crucial, for example, for the readout of a highly packed spin based quantum memory or the detection of unlabeled metalloproteins in a single cell.
In conventional EPR spectrometers, the energy transfer from the spins to the cavity at a Purcell enhanced rate \cite{Purcell1946} plays an essential role \cite{Schweiger2001, Bienfait2016a, Eichler2017}
and requires the spins to be resonant with the cavity, however the size of the cavity (limited by the wavelength) makes it difficult to improve the spatial resolution.
Here, we demonstrate a novel EPR spectrometer using a single artificial atom as a sensitive detector of spin magnetization.
The artificial atom, a superconducting flux qubit, provides advantages both in terms of its quantum properties and its much stronger coupling with magnetic fields.
We have achieved a sensitivity of $\sim$400 spins/$\sqrt{\mathrm{Hz}}$ with a magnetic sensing volume around 10$\mathbf{^{-14} \lambda^3}$ (50 femto-liters).
This corresponds to an improvement of two-order of magnitude in the magnetic sensing volume compared with the best cavity based spectrometers while maintaining a similar sensitivity as those spectrometers \cite{Bienfait2016, Probst2017a}.
Our artificial atom is suitable for scaling down and thus paves the way for measuring single spins on the nanometer scale.
}
 
EPR spectroscopy is an essential tool for characterizing the \green{properties of} electron spins in materials.
Due to the wide variety of EPR applications, significant efforts have been devoted to improving both its sensitivity and spatial resolution.
A conventional EPR spectrometer relies on energy exchange (transverse) coupling, where the spins and detector should be resonant.
In particular, in a leaky cavity limit, the spins mainly emits photons to the measurement chain at the Purcell enhanced relaxation rate \cite{Bienfait2016a}, and the detector absorbs the photon energy as a signal. 
Recently, sensitive EPR spectrometers based on a superconducting resonator have been realized \cite{Eichler2017, Bienfait2016, Bienfait2017, Probst2017a} with a measurement chain that uses a quantum limited amplifier.
This approach limits the size of the device according to the wavelength, and so such spectrometers may not scale well at a smaller size.
On the other hand, it is also possible to observe the EPR phenomenon without a cavity and magnetization detection \cite{Chamberlin1979} is one such example.
Magnetically induced force detection \cite{Rugar2004} has recently been demonstrated that achieves high sensitivity and spatial resolution.
In these cases, energy transfer between spins and the detector is suppressed due to the large detuning, thus the signal is detected without significant disturbance to the spin system.
However, such non-resonant methods still require improved in their sensitivity.

In this paper, we demonstrate sensitive local EPR spectroscopy using an artificial atom (a superconducting flux qubit \cite{Orlando1999}) as a magnetic field sensor \cite{Ilichev2007, Bal2012}.
The superconducting flux qubit has two distinct states corresponding to clockwise and anti-clockwise circulating currents $I_p$.
Such current states can be strongly coupled with magnetic fields induced by the spins.
The magnetic coupling causes the resonance frequency of the flux qubit to shift thus enabling EPR spectroscopy with little disturbance to the spin system.
The interaction strength induced by the persistent current states is much larger \cite{Marcos2010, Zhu2011, Saito2013} than that of resonator based systems \cite{Eichler2017, Bienfait2016,  Probst2017a, Bienfait2017, Kubo2010, Bushev2011}.
This interaction also has a smaller spin-to-device distance dependence than a spin-spin interaction, which enables us to prove distant spins with high sensitivity.
Thus, the superconducting flux qubit must be suitable for the detection of a small number spins.

\begin{figure*}[htbp]
\centering
\includegraphics[clip]{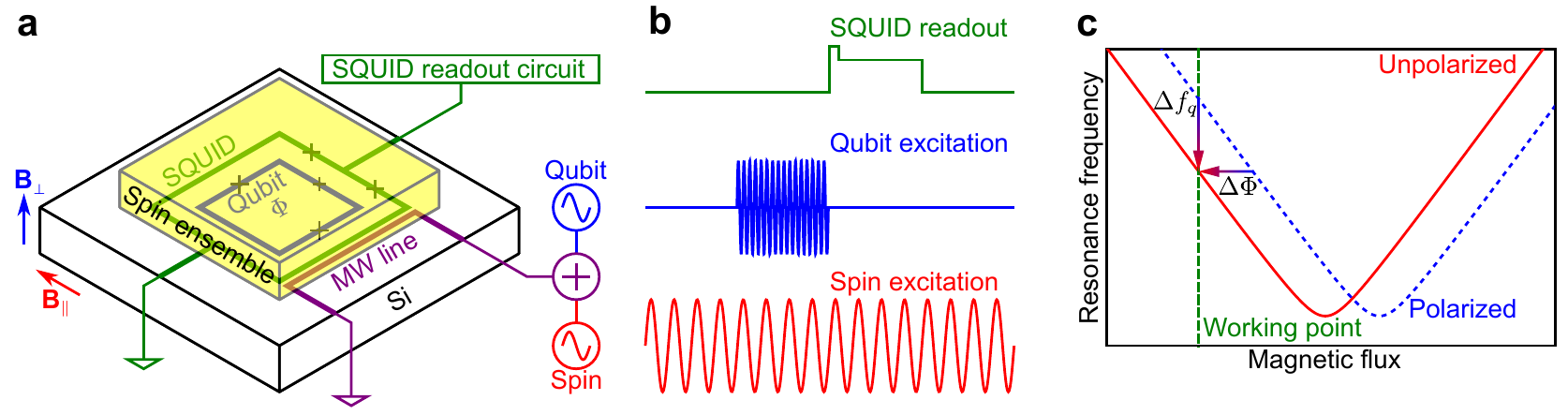}
\caption{
\textbf{Experimental setup for magnetic resonance spectroscopy using a superconducting flux qubit.} 
\textbf{a}, Experimental setup.
The spin ensemble is directly attached to the flux qubit chip.
The energy state of the flux qubit is read out by the SQUID.
The spin ensemble and the flux qubit are excited through the microwave (MW) line.
In-plane ($\Bin$, red arrow) and perpendicular ($\Bpp$, blue arrow) magnetic fields are applied by superconducting magnets to polarize the spin ensemble and to control the flux qubit. 
\textbf{b}, Pulse sequence used for the experiment.
SQUID readout pulses and qubit excitation pulses are used for the spectroscopy of the flux qubit.
In addition to these pulses, a continuous microwave excitation signal is applied to the same microwave line to excite the spin ensemble. 
\textbf{c}, Energy spectrum of the flux qubit. 
The resonance frequency is controlled by the magnetic flux penetrating the qubit loop.
For a fixed working point (green dashed line), the external magnetic flux $\Delta \Phi$ generated by the spin ensemble is detected from the change in the qubit resonance frequency $\Delta f_q$.
} 
\label{fig:1}
\end{figure*}

The principle of our approach to EPR spectroscopy is as follows.
We use a flux qubit to measure the magnetization of the spin (Fig. \ref{fig:1}a).
The resonance frequency of the flux qubit $f_q=\sqrt{\varepsilon(\Phi)^2 + \Delta^2}$ is sensitive to the magnetic flux penetrating the flux qubit loop $\Phi$, where $\varepsilon(\Phi) := 2I_p\left(\Phi-\Phi_0/2\right)/h$ is the frequency detuning, $\Phi_0$ is the magnetic flux quanta, $h$ is Planck's constant and $\Delta$ is the energy gap of the flux qubit.
Now spectroscopy of the flux qubit is performed by applying excitation and readout pulses to the device (Fig. \ref{fig:1}b), where the energy state of the flux qubit is read out by a superconducting quantum interference device (SQUID) using a switching method \cite{Wal2000} with 1000 repetitions.
The magnetic interaction between the spins and the flux qubit is realized by attaching the spin ensemble directly to the flux qubit chip (Fig. \ref{fig:1}a).
An additional magnetic flux $\Delta \Phi$ is generated by the attached spin ensemble, which in turn shifts the spectrum of the flux qubit.
Thus, when the working flux $\Phi$ is fixed, the spin polarization is detected as \green{a} resonance frequency shift $\Delta f_q$ (Fig. \ref{fig:1}c).
To perform EPR spectroscopy, we employ a continuous spin excitation signal, in addition to the microwave pulse for the flux qubit (Fig. \ref{fig:1}b).

Before performing EPR spectroscopy, we first characterize the flux qubit as a detector of magnetization from the spin ensemble (an Er$^{3+}$:Y$_2$SiO$_5$ crystal in this case).
By controlling the sample temperature $T$ and in-plane magnetic field $|\Bin|$, we can control the spin polarization ratio.
The signal $\Delta f_q$ can be used to calibrate the qubit based magnetometer.
Although we mainly apply in-plane magnetic field to the sample ($\left|\Bin\right| \gg \left|\Bpp\right|$), the spin ensemble generates perpendicular magnetization due to the anisotropic g-factor of the electron spins in the Er$^{3+}$:Y$_2$SiO$_5$ crystal \cite{Guillot-Noel2006, Sun2008, Budoyo2017a}.
\begin{figure*}[htbp]
\centering
\includegraphics[clip]{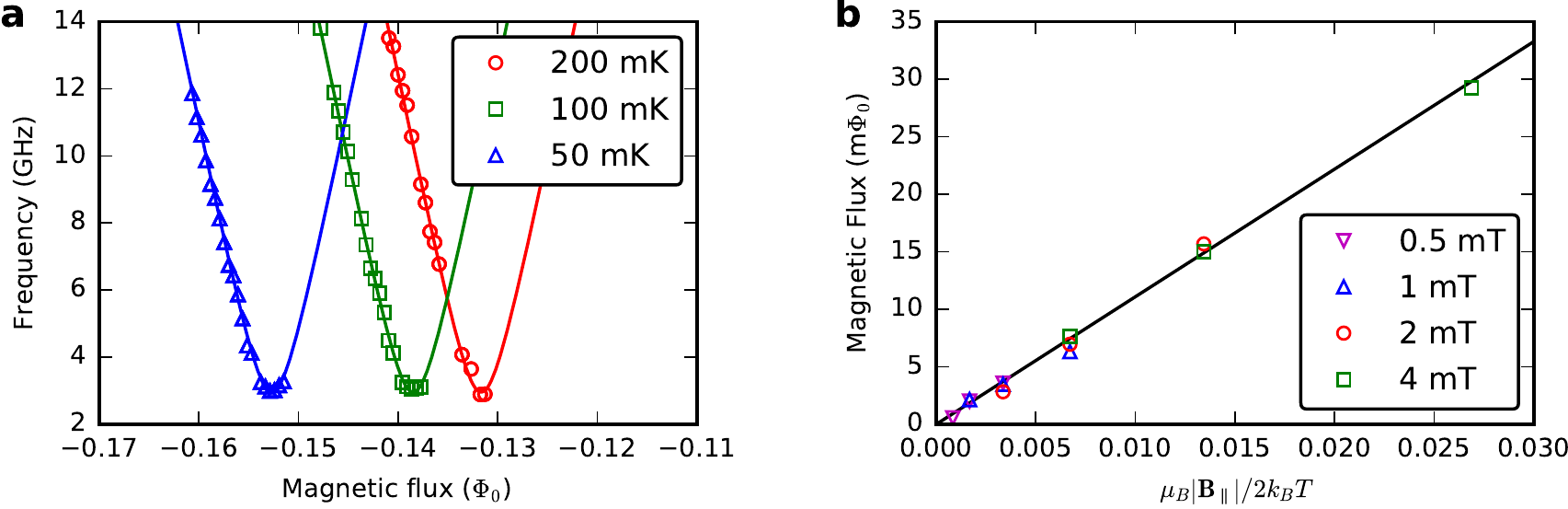}
\caption{\textbf{Detection of spin polarization with an Er$^{3+}$:Y$_2$SiO$_5$ sample.} \textbf{a}, Temperature dependence of the qubit spectrum under a 4 mT in-plane magnetic field. Open symbols are experimental data and solid lines are fitting. \textbf{b}, Magnetic flux detected by the shift of the qubit spectrum as a function of temperature and in-plane magnetic field. Open symbols are experimental data and solid lines are fitting.} 
\label{fig:2}
\end{figure*}
In Fig. \ref{fig:2}a, we plot the temperature dependence of the flux qubit spectrum under an in-plane magnetic field of 4 mT.
As the temperature increases, the flux qubit spectrum shifts to the positive flux side.
In Fig. \ref{fig:2}b, we summarize the in-plane magnetic field and temperature dependence of the   flux qubits' spectrum shift.
The linear fit reproduces the experimental results well. 
Although the entire magnetic field dependence is expected to be complicated due to the 7/2 nuclear spin of erbium atoms, our numerical simulation well reproduces the linear increase in the magnetization of our experimental setup as shown in Supplementary Information.

\begin{figure*}[htbp]
\centering
\includegraphics[clip]{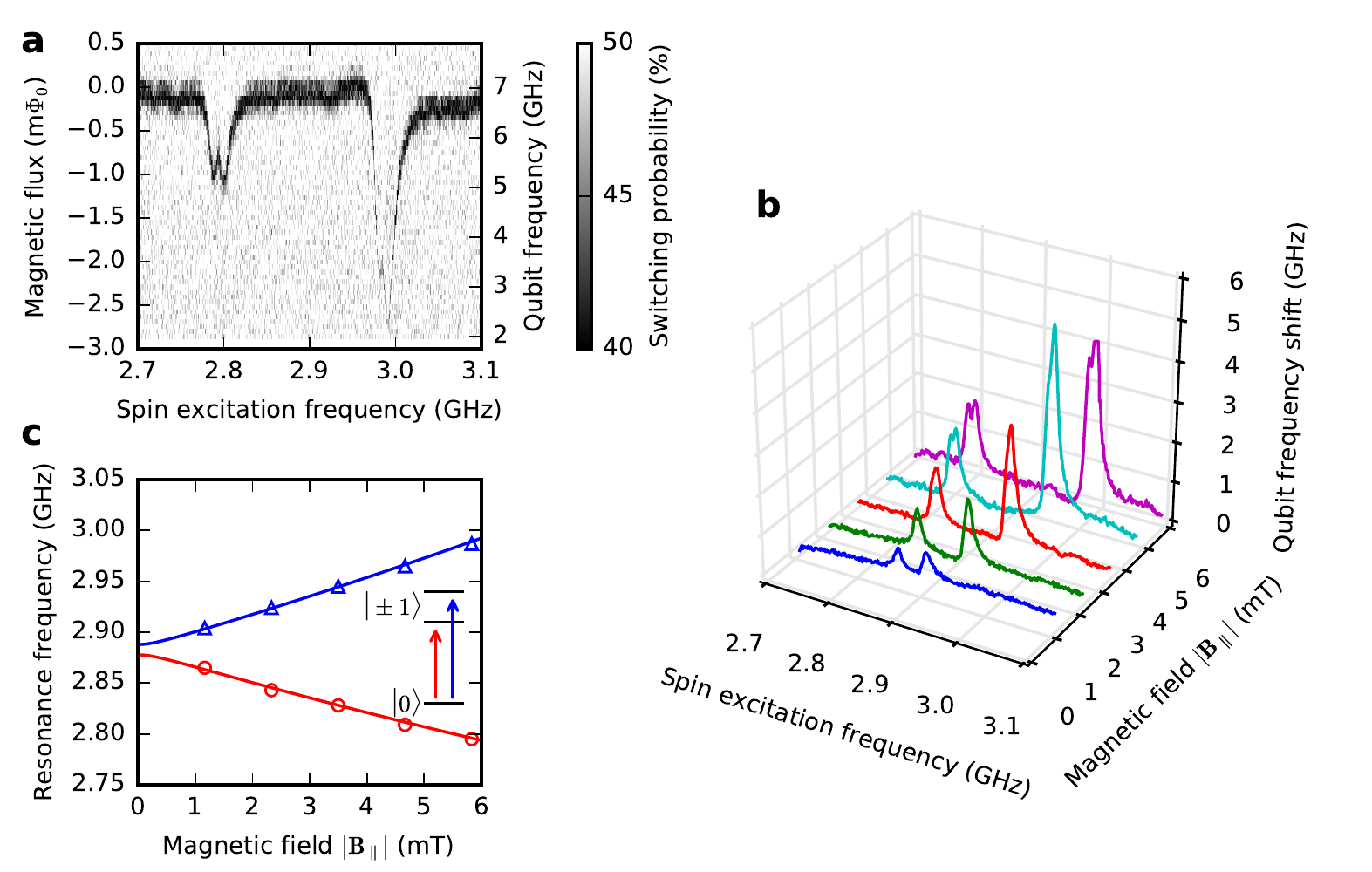}
\caption{\textbf{EPR spectroscopy.} \textbf{a}, 
 Results of EPR spectroscopy for NV centers in diamond under a 5.8 mT in-plane magnetic field. The qubit excitation frequency is converted to the corresponding magnetic flux. For the spectroscopy, the temperature is fixed at 20 mK to maximize the spin polarization ratio. \textbf{b}, EPR spectrum in several magnetic fields. The higher frequency peak in the 5.8 mT curve (magenta) has a smaller frequency shift due to the limitation of the dynamic range of the spectrometer. \textbf{c}, EPR peak frequency as a function of in-plane magnetic filed. The blue triangles (red circles) correspond to the transitions from the ground state to the first (second) excited state. The solid lines are the fitting curve calculated from the spin Hamiltonian. Inset: energy level structure of the NV center in diamond.}
\label{fig:3}
\end{figure*}

Next we performed an EPR experiment by exciting the spin ensemble using a microwave oscillating field. 
For this experiment, the nitrogen vacancy (NV) centers in diamond are employed as the characterized spin ensemble because its large zero-field splitting allows a high spin polarization ratio even under a small in-plane field.
The EPR spectrum, obtained under a 5.8 mT in-plane magnetic field with our continuous microwave spin excitation, is shown in Fig. \ref{fig:3}a.
Here the 2.88 GHz zero-field splitting in the spin one NV center ensures a large spin polarization ratio even in a small magnetic field regime.
For this experiment, the in-plane magnetic field is applied along [100] direction of the diamond crystal.
The bare resonance frequency of the flux qubit ($\sim$7 GHz) is detuned so that it is far from the expected resonance frequency of the NV centers ($\sim$3 GHz) by tuning the perpendicular magnetic field, $|\Bpp|$.
We observed that the frequency of the flux qubit decreases when we drive the spin with a frequency of $\sim$2.8 or $\sim$3.0 GHz.
Although an NV center has four possible orientation axes, every NV center is affected by the same amount of Zeeman splitting when the in-plane magnetic field is applied along the [100] direction. 
So the two observed resonances correspond to the transitions from the ground to the first and second excited states (see Fig. \ref{fig:3}c inset).
We also observe tiny splitting in each EPR peak, and this originates from a small misalignment ($\sim$ 3$^\circ$) of the magnetic field.
The different amplitudes of the two peaks are explained by considering the energy relaxation between three levels (see Supplementary Information).
We attribute the asymmetric lineshape of the resonance that we observed to the long energy relaxation time of the NV centers at low temperature \cite{Harrison2006, Amsuess2011}.
To obtain further insight into the EPR peaks, we perform EPR spectroscopy in various magnetic fields (Fig. \ref{fig:3}b).
\green{In Fig. \ref{fig:3}c (blue triangles and red circles), we plot the magnetic field $|\Bin|$ dependence of the EPR frequency.}
These experimental points are fitted with the transition frequency of the NV center calculated from the energy eigenvalues of the following spin Hamiltonian \cite{Loubser1978}:
\begin{equation}
	\hat{H}_s = g_e \mu_B\mathbf{B}\cdot\hat{\mathbf{S}} + hD\hat{S}_z^2 \red{ + hE\left( \hat{S}_y^2 - \hat{S}_x^2\right)},  
	\label{eq:1}
\end{equation}
where $g_e$ is the Land\'{e} g-factor,
\green{$\mu_B$ is the Bohr magneton,} $\mathbf{B}:=\Bin+\Bpp$ is the magnetic field,
$\hat{\mathbf{S}}=(\hat{S}_x, \hat{S}_y, \hat{S}_z)$ is the spin-one operator,
$D$ is the zero-field splitting, and $E$ is the strain.
Here, we assume a strain term of 5 MHz \cite{Saito2013}.
From the fitting constants, we derive $g_e$ and $D$ values of 2.05 and 2.883 GHz, respectively.
This result deviates slightly from the value reported in the literature \cite{Loubser1978} due to the magnetic field distortion near the superconductor caused by the Meissner effect.

The sensing volume of this spectrometer is estimated from the loop area and the effective thickness of the spin ensemble.
The loop area is the designed parameter of 47.2 $\mu$m$^2$.
Our effective thickness is defined as a typical length scale, in which the spin and the flux qubit interact strongly.
The interaction strength can be calculated numerically and the effective thickness is defined as $\sim$1 $\mu$m from calculated results for a flux qubit with a similar size to ours \cite{Marcos2010}.
By multiplying these values, the sensing volume is estimated to be $\sim$50 fL (5$\times$10$^{-17}$ m$^3$).
This value corresponds to a magnetic mode volume of $\sim$10$^{-14} \lambda^3$, and two orders of magnitude smaller than that obtained with a EPR spectrometer using a superconducting resonator \cite{Bienfait2016, Probst2017a}.

We \green{can} also estimate the minimum detectable number of spins per unit time.
\begin{figure*}[htbp]
\centering
\includegraphics[clip]{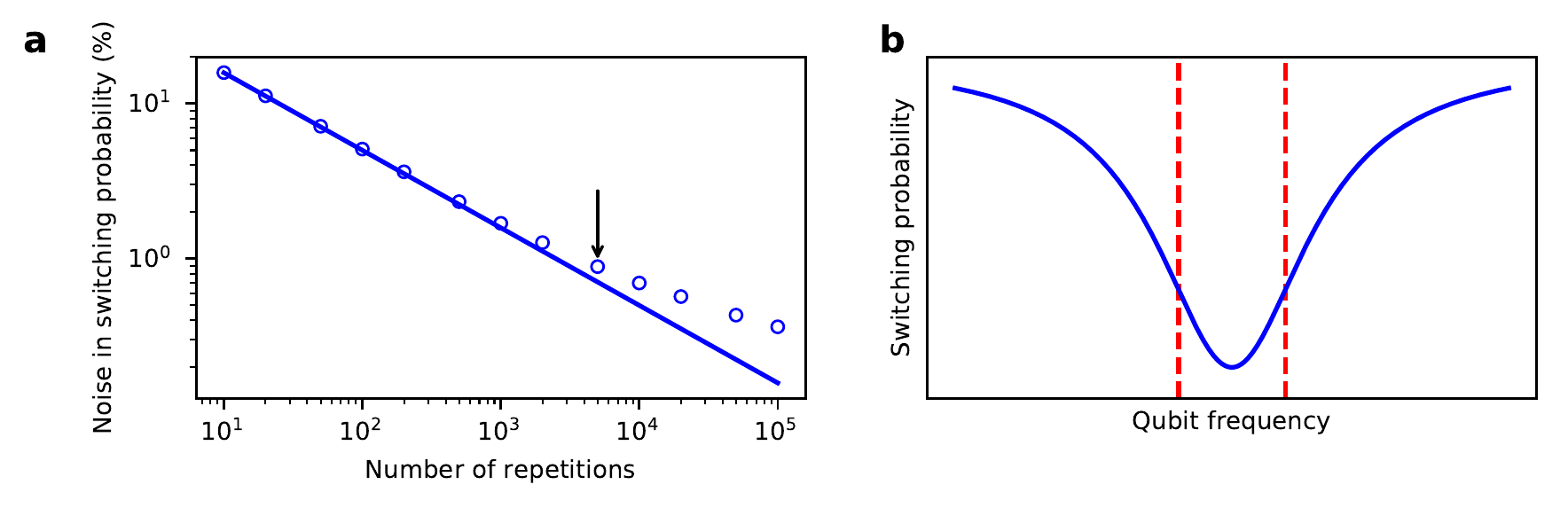}
\caption{\textbf{Estimation of sensitivity.} \textbf{a}, Noise in switching probability as a function of the number of repetitions. The solid line is the calculated $1/\sqrt{N_{\mathrm{rep}}}$ dependence. The arrow indicates the experimental data used to estimate the sensitivity. \textbf{b}, Conversion from switching probability noise to frequency noise. The most sensitive working point is indicated by the red dashed line. 
}
\label{fig:4}
\end{figure*}
For this purpose, we plot the measured noise in the switching probability as a function of the number of repetitions $N_{\mathrm{rep}}$ to obtain one experimental point (Fig. \ref{fig:4}a).
The noise does not follow the theoretical $1/\sqrt{N_{\mathrm{rep}}}$ scaling in \green{the} $N_{\mathrm{rep}} \gtrsim 2000$ region, possibly due to the slow drift of the system.
We use the noise for $N_{\mathrm{rep}}=5000$, which corresponds to integration for one second, to estimate the sensitivity per unit time in a real experimental environment.
\blue{By setting the working point of \green{our} flux qubit at the steepest point of the Lorentzian resonance peak, we obtain the best sensitivity (Fig. \ref{fig:4}b), and the noise in the switching probability is converted to frequency noise.}
Furthermore, we need to convert this noise to the corresponding number of spins using the experimental parameters (see Methods section).

The frequency noise can be converted to flux noise using the slope of the flux qubit spectrum (Fig. \ref{fig:1}c), \green{where} the flux noise is converted to \green{a} fluctuation in spin number using \green{the} generated flux per spin.
This value is estimated using SQUID magnetometry \cite{Toida2016}.
By combining these values, the sensitivity is estimated to be \green{530 $\pm$ 320} spins/$\sqrt{\mathrm{Hz}}$.
To check this approach we can also estimate the sensitivity using the following Hamiltonian, which represents the interaction between a single spin and a flux qubit:
\begin{equation}
	\hat{H} = h\left( \frac{f_q}{2} + \mathbf{g}\cdot\hat{\mathbf{S}}\right)\hat{\sigma}_z + \hat{H}_s.
\end{equation}
where $\mathbf{g}$ denotes the interaction strength (see Methods section).
Because the zero field splitting is much larger than the Zeeman energy in our experiment, the expectation values of $\mathbf{g}\cdot\hat{\mathbf{S}}$ for ground, and the first excited and second excited states are well approximated by 0, $-g_z$ and $g_z$, respectively. 
Thus, \green{the} frequency shift per single spin is $2g_z$.
Here, $g_z$ is estimated to be 4.4 kHz by using the Biot Savart law \cite{Zhu2011}.
Combining this value with the frequency noise, the sensitivity is estimated to be \green{300 $\pm$ 180} spins/$\sqrt{\mathrm{Hz}}$ (which is consistent with our original estimation).
\green{Such a sensitivity} is comparable to that of a resonator based EPR spectrometer with a quantum limited measurement chain \cite{Bienfait2016}. 

In summary, we have demonstrated highly sensitive micrometer-scale EPR spectroscopy using a superconducting flux qubit.
We estimate the sensitivity and the sensing volume of the spectrometer to be $\sim$400 spins/$\sqrt{\mathrm{Hz}}$ and $\sim$50 fL, respectively.
The sensitivity is comparable to that of EPR spectrometers using a superconducting resonator with a quantum limited amplifier, while the magnetic sensing volume is two orders of magnitude smaller than that of a resonator based spectrometer \cite{Bienfait2016, Probst2017a}.
A magnetic interaction between the qubit and the spin ensemble is realized without resonance between them, which is a completely different detection principle from that of the standard EPR spectrometer using transverse coupling.
As long as the change in the magnetization occurs, our local magnetic resonance scheme is applicable to any spin species including nuclear magnetic resonance.
In addition, it is possible to further reduce the sensing volume towards the realization of the nanoscale spectroscopy, because the size of the flux qubit loop is not limited by the wavelength.
Towards the detection of a single electron spin, a sensitivity improvement of three orders of magnitude is also possible by using a flux qubit with a narrower line width \cite{Stern2014, Yan2016}, by repeating the qubit measurement within a short period using a Josephson bifurcation amplifier \cite{Siddiqi2004} or with the dispersive readout method \cite{Blais2004}, 
and utilizing the quantumness of the qubit fully as discussed in the quantum sensing field \cite{Degen2017}.

\section{Methods}

\subsection{Experimental setup}
Magnetic flux generated by a spin ensemble is detected by a superconducting flux qubit with a loop area of 47.2 $\mu$m$^2$.
We used two spin ensembles for the experiment: 
10 ppm erbium doped Y$_2$SiO$_5$ single crystal (Scientific Materials, Inc.) is used for spin polarization detection and 
NV centers in type Ib diamond are used for EPR spectroscopy.
In-plane ($\Bin$) and perpendicular ($\Bpp$) magnetic fields are applied to the sample to polarize the spin ensemble and to control the flux qubit.
$\Bin$ and $\Bpp$ are parallel to the $\mathbf{D_1}$ and $\mathbf{D_2}$ axes of the Er$^{3+}$:Y$_2$SiO$_5$ crystal, respectively. 
The in-plane magnetic field $\Bin$ is oriented parallel to the [100] axis of the diamond crystal.
For the spectroscopy of the qubit and spin ensemble, a two-tone microwave signal is applied to the sample through an on-chip microwave line. 
The qubit state is read out by a SQUID with a repetition period of 200 $\mu$s and averaged over 1000 times.
All the measurements are performed in a dilution refrigerator, whose base temperature is lower than 20 mK.

\subsection{Derivation of system Hamiltonian in far detuned regime}
A single spin and flux qubit coupling system is described by the following Hamiltonian:
\begin{equation}
	\hat{H} = \frac{h\varepsilon}{2}\hat{s}_3 + \frac{h \Delta}{2}\hat{s}_1 + h \mathbf{g}\cdot\hat{\mathbf{S}}\hat{s}_3 + \hat{H}_s,
\end{equation}
where 
$\hat{s}_i$ is the Pauli matrix for the flux qubit,
$\mathbf{g} = (g_x, g_y, g_z)$ is the coupling strength between a spin and a flux qubit, 
$\hat{\mathbf{S}}$ is the spin operator vector associated with the spin, and
$\hat{H}_s$ is the spin Hamiltonian for the spin.
The axis dependence of $\mathbf{g}$ is attributed to the direction of the magnetic field generated by the flux qubit.
Here, we define the $z$ axis as the quantization axis of the spin.
By diagonalizing the flux qubit term, we obtain the following Hamiltonian:
\begin{equation}
	\hat{H} = \frac{hf_q}{2}\hat{\sigma}_3 + 
	h \mathbf{g}\cdot\hat{\mathbf{S}}\left(\hat{\sigma}_3\cos \theta + \hat{\sigma}_1 \sin \theta  \right)+ 
	\hat{H}_s,
\end{equation} 
where $\theta$ is the mixing angle defined by $\tan \theta = \Delta/\varepsilon$.
Because we operate the flux qubit far from the optimal point ($|\varepsilon| \gg \Delta$, $\sin \theta \sim 0$), we can safely neglect the transverse coupling term:
\begin{equation}
	\hat{H} = h\left( \frac{f_q}{2} + \mathbf{g}\cdot\hat{\mathbf{S}}\right)\hat{\sigma}_3 + \hat{H}_s.
	\label{eq:Hfinal}
\end{equation}
Thus, the resonance frequency of the flux qubit is modified by $\mathbf{g}\cdot\hat{\mathbf{S}}$ due to the interaction with a single spin.
In our EPR spectroscopy technique, we detect the difference between the qubit frequencies with and without spin resonance.
Without spin resonance, the qubit frequency is shifted due to the polarization of the spin.
On the other hand, the qubit frequency stays on the bare frequency when the spins resonate with the microwave, because time averaged polarization is zero thanks to the rotation of the spin vector.

\subsection{Estimation of sensitivity}
We estimated the sensitivity of this scheme as follows using experimental parameters.
The measured noise in the switching probability is converted to corresponding minimum detectable number of spins $N_{\mathrm{min}}$: 
\begin{equation}
N_{\mathrm{min}} = \delta P_e \left|\frac{\partial f_q}{\partial P_e}\right|\frac{\delta N}{\delta f_q},
\end{equation}
where $P_e$ and $\delta P_e$ are the switching probability and its noise per unit time, respectively, 
and ${\delta N}/{\delta f_q}$ is the frequency shift $\delta f_q$ of the flux qubit induced by $\delta N$ spins.
Here, we assume a Lorentzian lineshape for $P_e(f_q)$:
\begin{equation}
	P_e(f_q) = V\frac{\gamma_q^2}{(f_q - f_{q0})^2 + \gamma_q^2},
\end{equation}
where $V$ is the visibility of the readout, and $\gamma_q$ is the linewidth of the flux qubit (see Fig. \ref{fig:4}b).
We can easily derive the parameters $V$, $f_{q0}$ and $\gamma_q$ from the experiment.
To maximize the sensitivity, the excitation frequency $f_{q}$ is set at the steepest point of the $P_e(f_q)$ curve.
This condition is satisfied when $\left|f_q - f_{q0}\right|= \gamma_q / \sqrt{3}$ and the resulting slope $\left|{\partial P_e}/{\partial f_q}\right|$ is $3\sqrt{3}V/8\gamma_q$.
Thus, the sensitivity is expressed as follows:
\begin{equation}
	N_{\mathrm{min}} = \delta P_e \frac{8\gamma_q}{3\sqrt{3}V}\frac{\delta N}{\delta f_q}.
\end{equation}

There are two possible ways to derive ${\delta N}/{\delta f_q}$.
The first way is to decompose ${\delta N}/{\delta f_q}$ into ${\delta N}/{\delta \Phi}\cdot{\delta  \Phi}/{\delta f_q}$, where ${\delta f_q}/{\delta \Phi} = 2I_p/h$ corresponds to the persistent current of the flux qubit in a far detuned \blue{regime}.
Thus, the sensitivity is estimated by the following equation:
\begin{equation}
	N_{\mathrm{min}} = \delta P_e \frac{4h\gamma_q}{3\sqrt{3}VI_p}\frac{\delta N}{\delta \Phi}.
\end{equation}
${\delta \Phi}/{\delta N}$ is the generated magnetic flux per single spin and is estimated with another experiment, e.g. SQUID magnetometry \cite{Toida2016}.

We can also estimate the sensitivity using the system Hamiltonian [Eq. (\ref{eq:Hfinal})], because it gives the qubit frequency shift per single spin.
For example, we obtain
\begin{equation}
	N_{\mathrm{min}} = \delta P_e\frac{8\gamma_q}{3\sqrt{3}Vg_z}
\end{equation}
for a spin-half system by substituting $\delta N/\partial f_q$.

\section{Acknowledgments}
We thank N.~Mizuochi for characterizing the NV centers in diamond.
We also thank B.~Rangga and I.~Mahboob for helpful discussions. 
This work was supported by CREST, JST, by JSPS KAKENHI (Grant No. 15K17732), and in part by MEXT Grant-in-Aid for Scientific Research on Innovative Areas ``Science of hybrid quantum systems'' (Grant No. 15H05869 and 15H05870).

\section{Author contributions}
All the authors contributed extensively to the work presented in this paper.
H.T. carried out the measurements and data analysis.
X.Z. and S.S. designed and fabricated the flux qubit and associated devices while S.S. and
K.K. designed and developed the flux qubit measurement system.
Y.M. and W.J.M. provided theoretical support.
H.T. wrote the manuscript, with feedback from all the authors.


\begin{thebibliography}{10}

\bibitem{Schweiger2001}
A.~Schweiger and G.~Jeschke.
\newblock {\em Principles of pulse electron paramagnetic resonance}.
\newblock Oxford University Press,  (2001).

\bibitem{Purcell1946}
E.~M. Purcell, Phys. Rev.{ \bf 69}, 681 (1946).

\bibitem{Bienfait2016a}
A.~Bienfait, J.~J. Pla, Y.~Kubo, X.~Zhou, M.~Stern, C.~C. Lo, C.~D. Weis,
  T.~Schenkel, D.~Vion, D.~Esteve, J.~J.~L. Morton and P.~Bertet, Nature{ \bf
  531}, 74--77 (2016).

\bibitem{Eichler2017}
C.~Eichler, A.~J. Sigillito, S.~A. Lyon and J.~R. Petta, Phys. Rev. Lett.{ \bf
  118}, 037701 (2017).

\bibitem{Bienfait2016}
A.~Bienfait, J.~J. Pla, Y.~Kubo, M.~Stern, X.~Zhou, C.~C. Lo, W.~C.~D.,
  T.~Schenkel, M.~L.~W. Thewalt, D.~Vion, D.~Esteve, B.~Julsgaard,
  K.~M{\o}lmer, J.~J.~L. Morton and P.~Bertet, Nature Nanotechnology{ \bf 11},
  253--257 (2016).

\bibitem{Probst2017a}
S.~Probst, A.~Bienfait, P.~Campagne-Ibarcq, J.~J. Pla, B.~Albanese, J.~F. D.~S.
  Barbosa, T.~Schenkel, D.~Vion, D.~Esteve, K.~M{\o}lmer, J.~J.~L. Morton,
  R.~Heeres and P.~Bertet, Applied Physics Letters{ \bf 111}, 202604 (2017).

\bibitem{Bienfait2017}
A.~Bienfait, P.~Campagne-Ibarcq, A.~H. Kiilerich, X.~Zhou, S.~Probst, J.~J.
  Pla, T.~Schenkel, D.~Vion, D.~Esteve, J.~J.~L. Morton, K.~M{\o}lmer and
  P.~Bertet, Phys. Rev. X{ \bf 7}, 041011 (2017).

\bibitem{Chamberlin1979}
R.~Chamberlin, L.~Moberly and O.~Symko, Journal of Low Temperature Physics{ \bf
  35}, 337 (1979).

\bibitem{Rugar2004}
D.~Rugar, R.~Budakian, H.~J. Mamin and B.~W. Chui, Nature{ \bf 430}, 329--332
  (2004).

\bibitem{Orlando1999}
T.~P. Orlando, J.~E. Mooij, L.~Tian, C.~H. van~der Wal, L.~S. Levitov, S.~Lloyd
  and J.~J. Mazo, Phys. Rev. B{ \bf 60}, 15398--15413 (1999).

\bibitem{Ilichev2007}
E.~Il'ichev and Y.~S. Greenberg, EPL (Europhysics Letters){ \bf 77}, 58005
  (2007).

\bibitem{Bal2012}
M.~Bal, C.~Deng, J.-L. Orgiazzi, F.~Ong and A.~Lupascu, Nature Communications{
  \bf 3}, 1324 (2012).

\bibitem{Marcos2010}
D.~Marcos, M.~Wubs, J.~M. Taylor, R.~Aguado, M.~D. Lukin and A.~S. Sørensen,
  Phys. Rev. Lett.{ \bf 105}, 210501 (2010).

\bibitem{Zhu2011}
X.~Zhu, S.~Saito, A.~Kemp, K.~Kakuyanagi, S.~Karimoto, H.~Nakano, W.~J. Munro,
  Y.~Tokura, M.~S. Everitt, K.~Nemoto, M.~Kasu, N.~Mizuochi and K.~Semba,
  Nature{ \bf 478}, 221 (2011).

\bibitem{Saito2013}
S.~Saito, X.~Zhu, R.~Ams\"{u}ss, Y.~Matsuzaki, K.~Kakuyanagi, T.~Shimo-Oka,
  N.~Mizuochi, K.~Nemoto, W.~J. Munro and K.~Semba, Phys. Rev. Lett.{ \bf 111},
  107008 (2013).

\bibitem{Kubo2010}
Y.~Kubo, F.~R. Ong, P.~Bertet, D.~Vion, V.~Jacques, D.~Zheng, A.~Dréau, J.-F.
  Roch, A.~Auffeves, F.~Jelezko, J.~Wrachtrup, M.~F. Barthe, P.~Bergonzo and
  D.~Esteve, Phys. Rev. Lett.{ \bf 105}, 140502 (2010).

\bibitem{Bushev2011}
P.~Bushev, A.~K. Feofanov, H.~Rotzinger, I.~Protopopov, J.~H. Cole, C.~M.
  Wilson, G.~Fischer, A.~Lukashenko and A.~V. Ustinov, Phys. Rev. B{ \bf 84},
  060501 (2011).

\bibitem{Wal2000}
C.~H. van~der Wal, A.~C.~J. ter Haar, F.~K. Wilhelm, R.~N. Schouten, C.~J.
  P.~M. Harmans, T.~P. Orlando, S.~Lloyd and J.~E. Mooij, Science{ \bf 290},
  773--777 (2000).

\bibitem{Guillot-Noel2006}
O.~Guillot-No\"{e}l, P.~Goldner, Y.~L. Du, E.~Baldit, P.~Monnier and
  K.~Bencheikh, Phys. Rev. B{ \bf 74}, 214409 (2006).

\bibitem{Sun2008}
Y.~Sun, T.~B\"{o}ttger, C.~W. Thiel and R.~L. Cone, Phys. Rev. B{ \bf 77},
  085124 (2008).

\bibitem{Budoyo2017a}
R.~P. Budoyo, K.~Kakuyanagi, H.~Toida, Y.~Matsuzaki, W.~J. Munro, H.~Yamaguchi
  and S.~Saito, preprint arXiv:1710.10801v1 [cond-mat.supr-con]{ \bf }.

\bibitem{Harrison2006}
J.~Harrison, M.~Sellars and N.~Manson, Diamond and Related Materials{ \bf 15},
  586--588 (2006).

\bibitem{Amsuess2011}
R.~Ams\"uss, C.~Koller, T.~N\"obauer, S.~Putz, S.~Rotter, K.~Sandner,
  S.~Schneider, M.~Schramb\"ock, G.~Steinhauser, H.~Ritsch, J.~Schmiedmayer and
  J.~Majer, Phys. Rev. Lett.{ \bf 107}, 060502 (2011).

\bibitem{Loubser1978}
J.~H.~N. Loubser and J.~A. van Wyk, Reports on Progress in Physics{ \bf 41},
  1201 (1978).

\bibitem{Toida2016}
H.~Toida, Y.~Matsuzaki, K.~Kakuyanagi, X.~Zhu, W.~J. Munro, K.~Nemoto,
  H.~Yamaguchi and S.~Saito, Applied Physics Letters{ \bf 108}, 052601 (2016).

\bibitem{Stern2014}
M.~Stern, G.~Catelani, Y.~Kubo, C.~Grezes, A.~Bienfait, D.~Vion, D.~Esteve and
  P.~Bertet, Phys. Rev. Lett.{ \bf 113}, 123601 (2014).

\bibitem{Yan2016}
F.~Yan, S.~Gustavsson, A.~Kamal, J.~Birenbaum, A.~P. Sears, D.~Hover, T.~J.
  Gudmundsen, D.~Rosenberg, G.~Samach, S.~Weber, J.~L. Yoder, T.~P. Orlando,
  J.~Clarke, A.~J. Kerman and W.~D. Oliver, Nature Communications{ \bf 7},
  12964 (2016).

\bibitem{Siddiqi2004}
I.~Siddiqi, R.~Vijay, F.~Pierre, C.~M. Wilson, M.~Metcalfe, C.~Rigetti,
  L.~Frunzio and M.~H. Devoret, Phys. Rev. Lett.{ \bf 93}, 207002 (2004).

\bibitem{Blais2004}
A.~Blais, R.-S. Huang, A.~Wallraff, S.~M. Girvin and R.~J. Schoelkopf, Phys.
  Rev. A{ \bf 69}, 062320 (2004).

\bibitem{Degen2017}
C.~L. Degen, F.~Reinhard and P.~Cappellaro, Rev. Mod. Phys.{ \bf 89}, 035002
  (2017).

\end{thebibliography}

\end{document}